\documentclass[paper=a4, fontsize=11pt]{article}
\usepackage{iftex,amsmath}
\iftutex
\usepackage{unicode-math}
\else
\usepackage{amssymb}
\def\z"{}
\def\UnicodeMathSymbol#1#2#3#4{%
 \ifnum#1>"A0
   \DeclareUnicodeCharacter{\z#1}{#2}%
  \fi}
\input{unicode-math-table}

\fi
\usepackage{amsthm}
\usepackage{enumitem}
\usepackage{cancel}
\usepackage{appendix}
\usepackage{graphicx}
\usepackage{geometry}
\usepackage{url}
\usepackage[colorlinks=true,linkcolor=blue]{hyperref}
\usepackage{sectsty}
\usepackage{multirow}
\allsectionsfont{ \normalfont\scshape \textbf}
\newlist{steps}{enumerate}{1}
\setlist[steps, 1]{label = Step \arabic*:}
\title{Why is superconductivity absent in bulk Infinite-layer Nickelates?}
\author{Akariti Sharma$^{a}$, Bharathiganesh Devanarayanan $^{a,b,*}$, Pratik D.Patel$^{a}$,\\ Navinder Singh$^{a}$\\
        \small $^{a}$Theoretical Physics Division, Physical Research Laboratory, Navrangpura Ahmedabad, India - 380009 \\
        \small $^{b}$Indian Institute of Technology, Gandhinagar, Palaj, Gujarat, India - 382355 \\\\
        \small $^{*}$Corresponding author:\,Bharathiganesh Devanarayanan; \tt{dbharathiganesh@gmail.com}
}
\date{} %leave blank

\begin{document}

\maketitle

\begin{abstract}
We answer in affirmative the absence of superconductivity in bulk prepared samples of doped infinite-layer Nickelates through a simple magnetostatic model. In an independent approach we also employ Density Functional Theory (DFT) calculations for the same investigation. Our investigations reveal that it is the formation of Nickel clusters and Nickel deficient regions in bulk prepared samples (as opposed to thin films) that leads to the absence of superconductivity in bulk prepared samples.  
\noindent
\end{abstract}

Ever since the discovery of high temperature (T$_c$) superconductivity in doped Cuprates\cite{cuprate1}, there have been several layers of mysteries in understanding the mechanism of superconductivity in such materials. One approach to possibly understand the mechanism was to synthesise other families of high T$_c$ superconductors and then compare their properties with that of Cuprates. Ni$^{+1}$ with d$^9$ configuration similar to the configuration of Copper in Cuprates made Nickelates a leading candidate for such a family. Several decades of search for such a superconducting cousin finally found success when Li, Danfeng et al.\cite{li2019superconductivity} synthesised superconducting thin films of NdNiO$_{2}$ doped with Sr (Nd$_{1-x}$Sr$_{x}$NiO$_2$). \\
Contrary to just helping our understanding of superconducting mechanism in Cuprates, Nickelate superconductors came with their own additional layers of mysteries\cite{singh2019road}. Li, Qing et al.\cite{li2020absence} had synthesised bulk samples of Nd$_{1-x}$Sr$_{x}$NiO$_2$ and have reported absence of superconductivity in the same. Several other groups had synthesised bulk samples of infinite layer Nickelates and superconductivity has not been seen in any of them\cite{Nickelateabsence} but it is observed in thin films of several nanometer thickness \cite{lee2020aspects}. It was thought that the substrate on which the thin film is mounted is playing a key role in emergence of superconductivity. This also lead to prepositions that superconductivity in the thin films could be an interface phenomena in the interface of the material Nd$_{1-x}$Sr$_{x}$NiO$_2$ and the substrate SrTiO$_{3}$.\\
Here we emphasise that one should clearly understand that the substrate can either play an indirect role by providing strain modulation and crystal stability to the thin film sample or directly participate in the superconducting mechanism. It should be made clear that only the latter makes superconductivity in Nickelates an interfacial phenomena and not the former one. Experiments on Meissner effects confirms that the substrate does not directly participate in superconducting mechanism ruling out that it is an interface phenomena\cite{frontier}.  \\
\begin{figure}[!h]
    \centering
    \includegraphics[scale = 0.25]{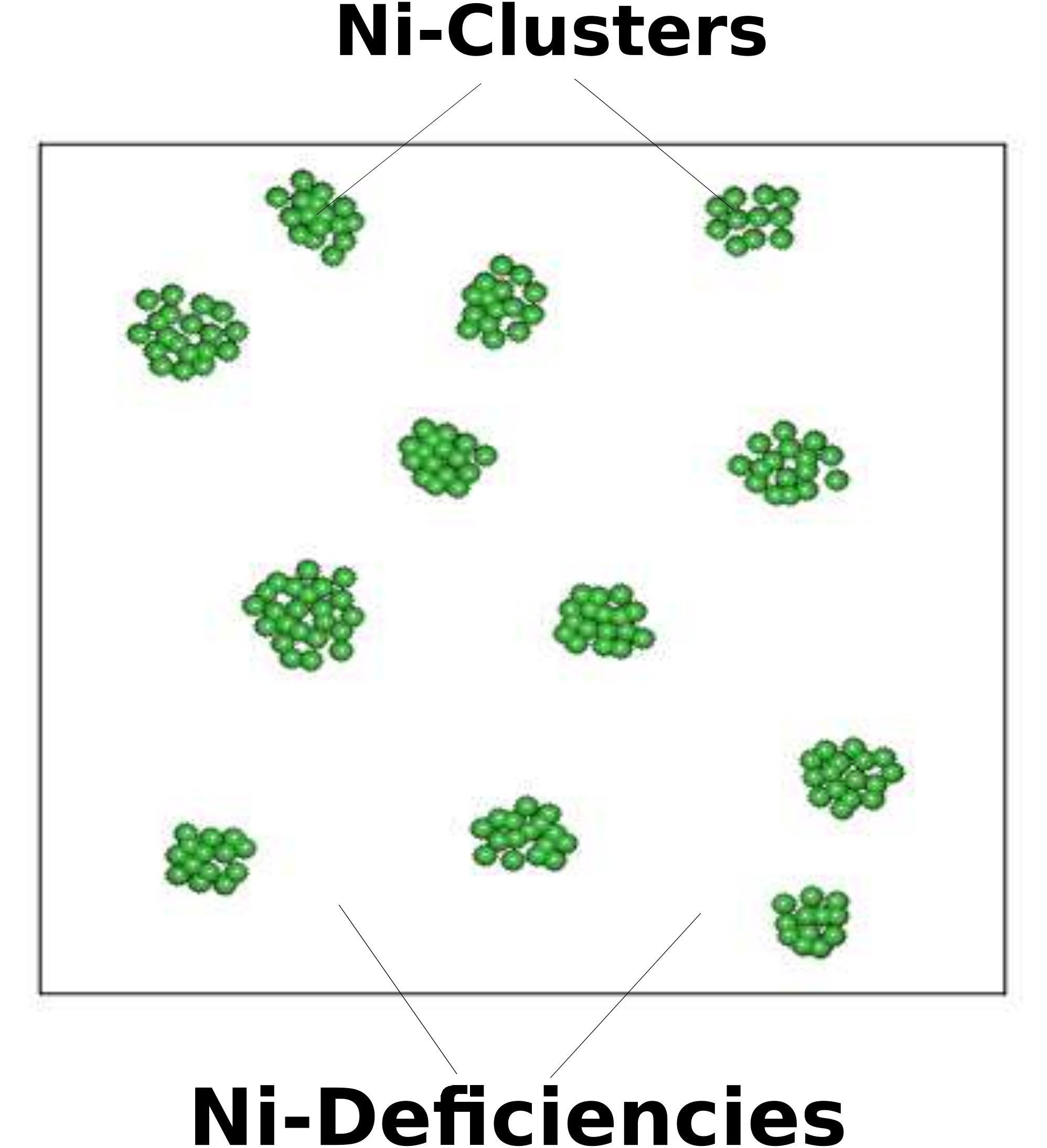}
    \caption{Schematic diagram of Ni-clusters in NdNiO$_2$ bulk sample. }
    \label{fig2}
\end{figure}
\textbf{It has been reported that by Li, Qing et al.\cite{li2020absence} that in the bulk samples of infinite layer Nickelates there are Ni deficiencies and these Ni atoms form clusters and remain together with the grains of Nd$_{1-x}$Sr$_{x}$NiO$_2$}(fig.\ref{fig2}). It has also been mentioned that the size of such clusters can go up to a few $\mu$ms [private communication]. Since Ni metal is a ferromagnet these Nickel clusters of micro and sub-micro meter size can also be considered as tiny ferromagnets present within the sample.  It should also be noted that the first critical field ($H_{c_{1}}$) of Nd$_{1-x}$Sr$_{x}$NiO$_2$ in thin film samples is reported to be 0.79 mT. \\
\textbf{In our simple model we estimate the magnetic fields produced by these Nickel clusters and notice that it can be greater than the first critical field $H_{c_{1}}$}. The magnetic fields produced by the Nickel clusters can be calculated either from the magnetic dipole model (point model) or the Amp\'erian loop model (extended model). In the first model the Ni cluster is modeled to be magnetic dipole with a moment equal to the sum of individual moments from each Nickel atom. This is very well justified because the temperatures are well below the ordering temperature of Nickel and the moments are definitely aligned. The magnetic field due to a magnetic dipole is given by the formula:
\begin{equation}\label{dipole}
    \mathbf{B}(\mathbf{r}) = \frac{\mu_{0}}{4\pi} \left(\frac{3 \mathbf{\hat{r}}\left(\mathbf{\hat{r}.m}\right)-\mathbf{m}}{|\mathbf{r}|^{3}}\right) 
\end{equation}
where $\mathbf{m}$ is the total magnetic dipole moment of the cluster and $\mathbf{r}$ is the vector pointing from the center of the magnetic dipole to the point where the magnetic field is measured.\\
In the Amp\'erian loop model the nickel clusters are modelled as tiny current carrying loops with the radius being the size of the Nickel cluster and the magnetic moment the sum of magnetic moments all Nickel atoms in the cluster (see fig.\ref{fig1}). The current in the fictitious loop can then be straight forwardly calculated as:
\begin{equation}
    \mathbf{m} = \mathbf{IA},
\end{equation}
Here $\mathbf{I}$ is the current in the loop and \textbf{A} is the area of the loop.
\begin{figure}[h]
    \centering
    \includegraphics[scale = 0.25]{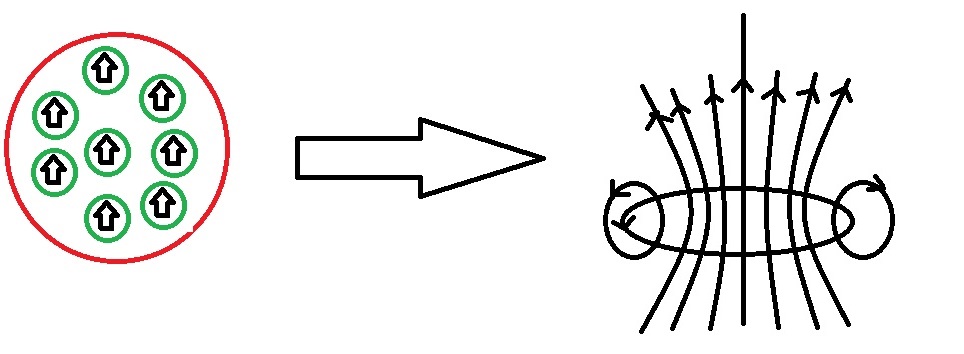}
    \caption{Nickel cluster modeled as a current carrying loop of same radius in the Amp\'erian model. }
    \label{fig1}
\end{figure}
To get a rough estimate of the total magnetic moment of the Nickel cluster we need to approximately calculate the number of Nickel atoms in each cluster. Suppose the radius of a Nickel cluster is r (1$\mu$m). Given the lattice constant of Ni metal is $a$ ( $3.5\times10^{-10}$ m) and Nickel crystallises in a Face centered Cubic (FCC) lattice, the number of Nickel atoms in each cluster is approximately:
\begin{equation}
    N_c = \frac{4\frac{4\pi}{3}(r)^{3}}{(a)^{3}} = \frac{4\frac{4\pi}{3}(10^{-6})^{3}}{(3.5\times10^{-10})^{3}} = 1.19\times10^{12},
\end{equation}
Since each Nickel atom contributes 3 Bohr magneton($\mu_B$) of magnetic moment, the total magnetic moment of Nickel cluster is:
\begin{equation}
    \mathbf{m} = 3\,N_C\,\mathbf{\mu_{B}}.
\end{equation}
From the value of the magnetic moment calculated above the magnetic field can be directly calculated from eqn.\eqref{dipole} in the magnetic dipole model. For the Amp\'erian loop model we will have to calculate current from the total magnetic moment. The magnetic field at a point in the axis of a current carrying loop is given by the following formula:
\begin{equation}
    \mathbf{B}_{axial}(\mathbf{x}) = \frac{\mu_{0} I}{2}\frac{r^{2}}{(r^{2}+x^{2})^{3/2}}\mathbf{\hat{z}} = \mu_{0}\mu_{B} \left(\frac{2r}{a\sqrt{r^{2}+x^{2}}}\right)^{3}\mathbf{\hat{z}},
\end{equation}
where $r$ is the radius of the current loop, $x$ is the distance from the centre of the loop to a point in the axis where magnetic field is calculated, $I$ is the current in the loop and $\mathbf{\hat{z}}$ is an unit vector along the axis and pointing away from the loop.\\
\begin{table}[]
\begin{center}
\begin{tabular}{ |c|c|c|c| } 
\hline
Radius of the cluster ($\mu$ms)  & Distance from the cluster ($\mu$ms) & Magnetic field (mT) \\
\hline
\multirow{3}{4em}{1} & 1 & 768.47 \\  
 & 5 & 16.39 \\ 
 & 10 & 2.14 \\ 
\hline
\multirow{3}{4em}{5} & 5 & 768.47 \\ 
 & 10 & 194.41 \\ 
 & 20 & 31.01 \\ 
\hline
\end{tabular}
\caption{\label{table}Magnetic field produced by the Nickel clusters of various sizes at various distances from the Nickel clusters.}
\end{center}
\end{table}
Our results for the magnetic field at various distances for various size of clusters have been tabulated in Table \ref{table}. The value for magnetic field obtained from both the models are of same order. We can observe the general trend that for distance up to an order higher than the radius of the cluster the field is of the order of mT. We have already noted that the size of these clusters are of the order of 1-6 $\mu$ms and their average inter cluster separation is of the order of 10 $\mu$ms.\\
\textbf{This leads to the conclusion that a magnetic field of an order of mT pervades the entire sample}.This magnetic field is of an order of magnitude higher than the $H_{c_{1}} (= 79\,$ Oe) of Nickelates and is enough to destroy the superconductivity.\\
\begin{figure}[h]
    \centering
    \includegraphics[scale = 0.45]{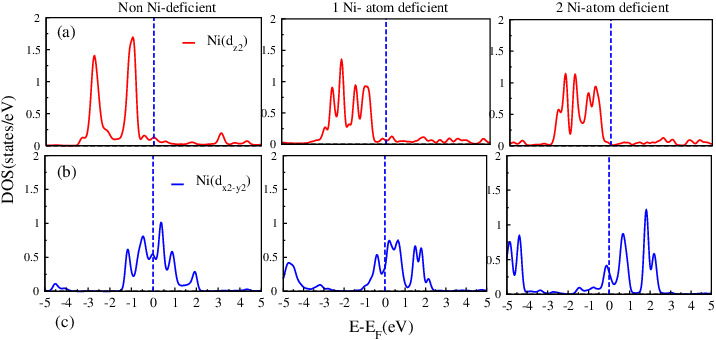}\\
    \includegraphics[scale = 0.35]{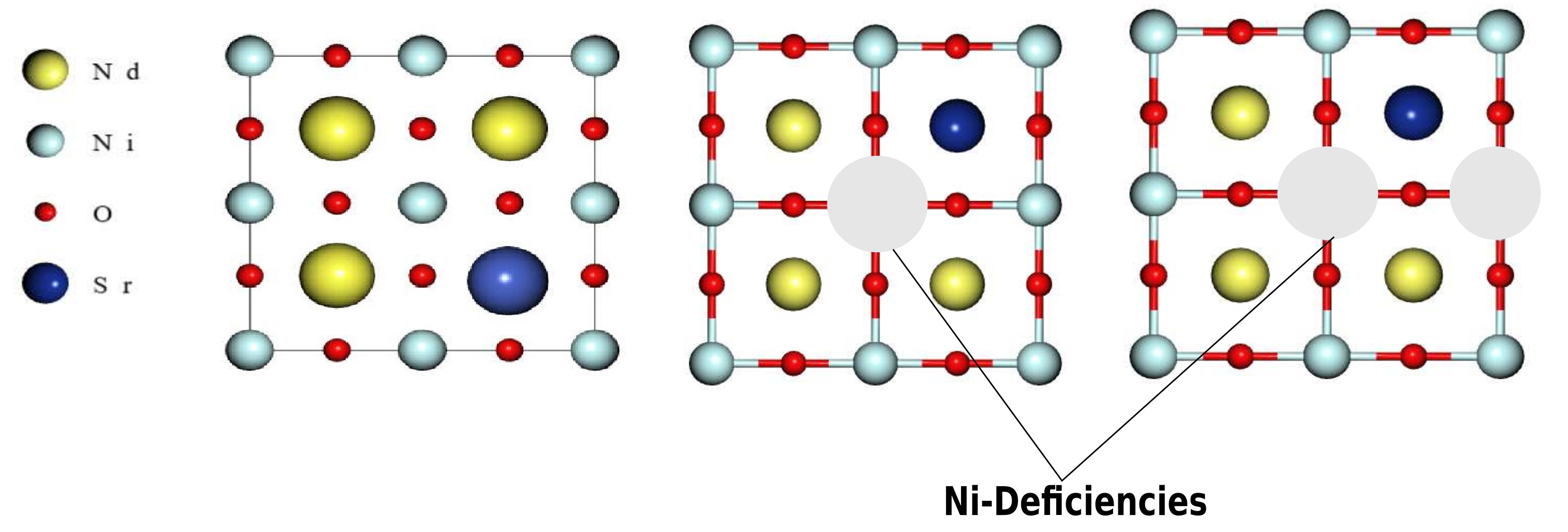}
    \caption{Orbital resolved PDOS for non Ni-deficient Nd$_{1-x}$Sr$_x$NiO$_2$ (0.25) in panel (a), one and two Ni-atom deficient supercells in panel (b) and (c), respectively.}
    \label{fig3}
\end{figure}
Another argument (from Nickel deficient regions) can be made in the following way. From our DFT calculations we infer that Nickel deficient regions modify the electronic structure. The DFT calculations have been performed for the system Nd$_{0.75}$Sr$_{0.25}$NiO$_2$ and the results are shown in fig.\ref{fig3}.  Nickel deficiency leads to the reduced weight of the critical Ni 3d$_{x^{2}-y^{2}}$ orbital at the Fermi level which is detrimental to superconductivity.\\
Therefore we conclude that the lack of superconductivity can be understood in the following way: The Nickel deficiency in the Nd$_{1-x}$Sr$_{x}$NiO$_2$ grains leads to decrease in the weight of Ni 3d$_{x^{2}-y^{2}}$ orbital which in turn affects superconductivity. Further the Nickel atoms form disconnected clusters between these grains. These clusters produce a magnetic field of an order of at least a milliTesla (mT) killing superconductivity if it exists. We have given two possible explanations for \textbf{"Why is superconductivity absent in bulk Infinite-layer Nickelates?"} that have been synthesised with the currently available state of the art chemical methods. It would be interesting to see in future the possibility of existence of superconductivity in bulk prepared Nickelate samples, if the two differences of Nickel deficiency and Nickel clusters between the film and bulk samples are resolved with more advanced chemical methods of synthesis.\\\\   

\vspace{1 cm}
{\bf{Acknowledgment}}\\\\
We thank Dr Qing Li for his correspondence regarding size of Nickel clusters.\\
Computation work related to these investigations were performed using the HPC resources (Vikram-100 HPC) project at Physical Research Laboratory (PRL).
\bibliographystyle{unsrt}
\bibliography{bib}
\end{document}